\documentclass[11pt,a4paper,fleqn]{article}
\usepackage[T1]{fontenc}
\usepackage{authblk}
\usepackage{amsmath,amssymb}
\usepackage{graphicx}
\usepackage[margin=0.5in]{geometry}
\usepackage{indentfirst}
\usepackage [autostyle, english = american]{csquotes}
\usepackage{setspace}
\usepackage{lineno}
\usepackage{enumerate}
\usepackage{enumitem}
\usepackage{changes}
\usepackage{parskip}
\MakeOuterQuote{"}

\title{FMRI Clustering and False Positive Rates\\
\large A {\it Letter} accepted by PNAS} 

\author[a]{Robert W. Cox\thanks{Corresponding author. E-mail address:
    robertcox@mail.nih.gov}} 
\author[a]{Gang Chen} 
\author[a]{Daniel R. Glen}
\author[a]{Richard C. Reynolds}
\author[a]{Paul A. Taylor}

\affil[a]{Scientific and Statistical Computing Core, National
  Institute of Mental Health, National Institutes of Health,
  Department of Health and Human Services, USA} \date{}
\begin{document}
\maketitle 

\noindent Recently, Eklund et al. (2016) analyzed clustering methods in standard
FMRI packages: AFNI (which we maintain), FSL, and SPM
\cite{Eklund2016}.  They claimed: 1) false positive rates (FPRs) in
traditional approaches are greatly inflated, questioning the validity
of ``countless published fMRI studies''; 2) nonparametric methods
produce valid, but slightly conservative, FPRs; 3) a common flawed
assumption is that the spatial autocorrelation function (ACF) of FMRI
noise is Gaussian-shaped; and 4) a 15-year-old bug in AFNI's
3dClustSim significantly contributed to producing ``particularly
high'' FPRs compared to other software. We repeated simulations from
\cite{Eklund2016} (Beijing-Zang data \cite{Biswal2010}, see
\cite{Cox065862}), and comment on each point briefly.

\subsection*{AFNI and 3dClustSim}  
Fig. \ref{fig:fig1}A-D compares results of the ``buggy'' and ``fixed''
3dClustSim. For each simulation, the typical difference was small:
$\Delta{\rm FPR} \lesssim3-5\%$ at per-voxel $p=0.01$ and
  $\lesssim1-2\%$ for $p=0.001$. The bug had only a minor impact.

Figs. 1-2 of \cite{Eklund2016} actually show similar FPRs for AFNI,
FSL-OLS, and SPM: most tests were in a range of $20-40\%$ FPR at
$p=0.01$ and $5-15\%$ FPR at $p=0.001$. (Nor did their famous 70\% FPR
come from AFNI.) Their Results' data simply do not support the
Discussion's statement that AFNI had ``particularly high'' FPRs.

\subsection*{Smoothness}  
To test the effect of assuming a Gaussian ACF in FMRI noise, an
empirical ``mixed ACF'' allowing for longer tails was computed from
residuals \cite{Cox065862}.  All FPRs (Fig. \ref{fig:fig1}E-F)
decreased.  Block designs remained $>5\%$, likely reflecting
dependence of the noise's spatial smoothness on temporal
frequency. Heavy-tails in spatial smoothness indeed have significant
consequences for clustering.

\subsection*{Nonparametric approach}  
A spatial model-free, nonparametric randomization approach was added
to AFNI's group-level GLM program, 3dttest++ \cite{Cox065862}. All
FPRs (Fig. \ref{fig:fig1}G-H) were within the nominal confidence
interval. While this approach shows promise (as in \cite{Eklund2016}),
it may not be feasible to generalize nonparametric permutations to
complicated covariate structures and models (e.g., complex ANOVA,
ANCOVA or LME) \cite{Chen2014,Chen2013}. 

\subsection*{Inflated FPRs}  
Several cases showed significant FPR inflation across existing FMRI
software within \cite{Eklund2016}'s testing framework. However,
deviations from nominal FPR were not uniformly large and depended
strongly on several factors.  Fig. \ref{fig:fig1} here and Fig. 1 of
\cite{Eklund2016} show quite good cluster results for stricter
per-voxel $p$-values (which \cite{Carp2012} found to be predominantly
used in FMRI analyses) and for event-related stimuli (emphasizing the
importance of good experimental design): FPR inflation was often
$\lesssim10\%$ (Beijing) or $\lesssim5\%$ (Cambridge), affecting only
clusters with marginally significant volume. 

We strongly disagree with \cite{Eklund2016}'s summary statement,
``Alarmingly, the parametric methods can give a very high degree of
false positives (up to $70\%$, compared with the nominal $5\%$).'' For
comparison, their own nonparametric method's results actually showed
up to 40\% FPR.  When characterizing results, medians or percentile
ranges are generally more informative summary statistics than
maxima. Looking backward, the typical ranges show {\it much} smaller
FPR inflation than what had been highlighted, and looking forward,
they provide useful suggestions for experimental design and analyses
(lower voxelwise $p$, event-related paradigms, etc.).  By
concentrating on the highest observed FPRs, the conclusions of Eklund
et al. were unnecessarily alarmist.

\subsection*{AFNI and 3dClustSim}  
The research and writing of the paper were supported by the NIMH and
NINDS Intramural Research Programs (ZICMH002888) of the NIH/DHHS,
USA. This work utilized the computational resources of the NIH HPC
Biowulf cluster (http://hpc.nih.gov).

\end{document}